\documentclass[12pt,preprint]{emulateapj}

\shortauthors{ENOTO ET AL.}
\shorttitle{SUZAKU OBSERVATION OF SGR~0501+4516 IN OUTBURST}

\begin{document}
\title{SUZAKU OBSERVATION OF THE NEW SOFT GAMMA REPEATER SGR~0501+4516 IN OUTBURST}

\author{T. Enoto\altaffilmark{1}, Y.E. Nakagawa\altaffilmark{2}, N. Rea\altaffilmark{3},
P. Esposito\altaffilmark{4,5}, D. G\"otz\altaffilmark{6}, K. Hurley\altaffilmark{7},
G.L. Israel\altaffilmark{8}, M. Kokubun\altaffilmark{9},\\
K. Makishima\altaffilmark{1,2},  S. Mereghetti\altaffilmark{4},
H. Murakami\altaffilmark{10}, K. Nakazawa\altaffilmark{1},
T. Sakamoto\altaffilmark{11}, L. Stella\altaffilmark{8}, \\
A. Tiengo\altaffilmark{4}, R. Turolla\altaffilmark{12,13},
S. Yamada\altaffilmark{1}, K. Yamaoka\altaffilmark{14},
A. Yoshida\altaffilmark{14}, and S. Zane\altaffilmark{13}
}
\altaffiltext{1}{Department of Physics, University of Tokyo,
    7-3-1 Hongo, Bunkyo-ku, Tokyo, 113-0033, Japan }
\altaffiltext{2}{Cosmic Radiation Laboratory,
  Institute of Physical and Chemical Research (RIKEN),
    Wako, Saitama, 351-0198, Japan}
\altaffiltext{3}{Astronomical Institute ``Anton Pannekoek",
University of Amsterdam, Kruislaan 403, 1098 SJ Amsterdam, The Netherlands}
\altaffiltext{4}{INAF--Istituto di Astrofisica Spaziale e Fisica Cosmica - Milano,
via E. Bassini 15, 20133 Milano, Italy}
\altaffiltext{5}{INFN-Istituto Nazionale di Fisica Nucleare, Sezione di Pavia, via A. Bassi 6, 27100 Pavia, Italy}
\altaffiltext{6}{CEA Saclay, DSM/Irfu/Service d'Astrophysique, Orme des Merisiers,
Bat. 709, 91191 Gif sur Yvette, France}
\altaffiltext{7}{Space Sciences Laboratory, 7 Gauss Way,
   University of California, Berkeley, CA 94720-7450, U.S.A.}
\altaffiltext{8}{INAF--Astronomical Observatory of Rome, via
Frascati 33, 00040, Monteporzio Catone (RM), Italy}
\altaffiltext{9}{Institute of Space and Astronautical Science, JAXA,
    3-1-1 Yoshinodai, Sagamihara, Kanagawa, Japan 229-8510}
\altaffiltext{10}{Department of Physics, Rikkyo University,
    3-34-1 Nishi-Ikebukuro, Toshima-ku, Tokyo, 171-8501, Japan }
\altaffiltext{11}{NASA Goddard Space Flight Center, Greenbelt, MD 20771, U.S.A.}
\altaffiltext{12}{University of Padua, Department of Physics , via
Marzolo 8, 35131 Padova, Italy}
\altaffiltext{13}{MSSL, University
College London, Holmbury St. Mary, Dorking Surrey, RH5 6NT, UK}
\altaffiltext{14}{Department of Physics \& Mathematics, Aoyama Gakuin University,
    Sagamihara, Kanagawa, 229-8558, Japan }

\begin{abstract}

We present the first {\it Suzaku} observation of the new Soft Gamma Repeater
SGR\,0501+4516, performed on 2008 August 26, 
four days after the onset of bursting activity 
of this new member of the magnetar family.
The soft X-ray persistent emission was detected 
with the X-ray Imaging Spectrometer (XIS) at a 0.5--10 keV flux of
$3.8\times10^{-11}$\,erg~s$^{-1}$cm$^{-2}$, 
 with a spectrum well fitted
 by an absorbed blackbody plus power-law model. 
The source pulsation was confirmed at a period of $5.762072\pm0.000002$~s,
and 32 X-ray bursts were detected by the XIS, 
four of which were also detected at higher energies 
by the Hard X-ray Detector (HXD).  
The strongest burst, which occurred at 03:16:16.9 (UTC), 
was so bright that it caused instrumental saturation, 
but its precursor phase, lasting for about 200\,ms, 
was detected successfully over the 0.5--200 keV range, 
with a fluence of $\sim 2.1 \times10^{-7}$\,erg\,cm$^{-2}$ 
and a peak intensity of about 89 Crab. 
The entire burst fluence is estimated to be $\sim 50 $ times higher.  
The precursor spectrum was very hard, 
and well modeled by a combination of two blackbodies.  
We discuss the bursting activity and X/$\gamma$-ray properties 
of this newly discovered Soft Gamma Repeater
in comparison with other members of the class.

\end{abstract}


\section{INTRODUCTION}

A new Soft Gamma Repeater (SGR), 
SGR 0501+4516, was discovered on 2008 August 22
by the {\em Swift} Burst Alert Telescope,
thanks to the detection of many short bursts
(Holland and Sato 2008; Barthelmy et al. 2008). 
Archival {\it ROSAT} data showed a faint X-ray
source consistent with the position of this new SGR, 
probably its quiescent counterpart (Kennea \& Mangano 2008).

SGRs are a sub-class of the so called ``magnetars", 
believed to be isolated neutron stars with very strong 
magnetic fields ($10^{14-15}$\,G: Thompson \& Duncan 1995). 
This extreme magnetic field powers their bright emission, 
rather than accretion or rotational power as  for most of the X-ray pulsars. 
Only four confirmed SGRs are known to date, 
all sharing common properties with the other magnetars 
(the Anomalous X-ray Pulsars, aka AXPs) such as 
i) a spin period in a very small range of values ($2-12$\,s), 
ii) large period derivatives ($10^{-13}-10^{-11}$\,s\,s$^{-1}$), 
iii) bright persistent X-ray emission ($10^{33-36}$\,erg\,s$^{-1}$),
compared to the rather dim infrared counterparts,
iv)  unpredictable bursting and flaring activity, 
with a large range of energetics ($10^{37-46}$\,erg\,s$^{-1}$) 
and timescales (ms to years), 
and v) transient radio emission (sometimes pulsed) connected with
their X-ray activity (Woods \& Thompson 2006; Mereghetti 2008).

We report in this {\it Letter} on a {\it Suzaku} Target of Opportunity (ToO) observation
of SGR\,0501+4516, the first new SGR discovered in our Galaxy in the last ten years, 
which was performed only four days after the
bursting activation.


\begin{figure*}[htbp]
\begin{center}
\includegraphics[scale=0.37]{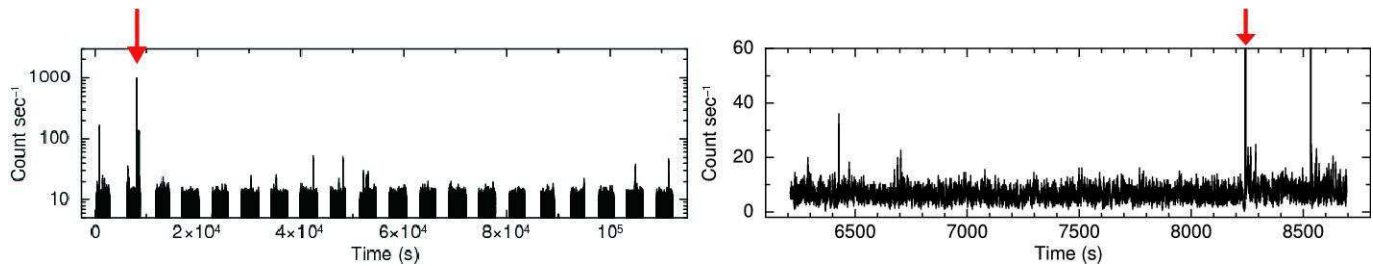}
\caption{{\em Left}:
2-s binned 0.4--10~keV light-curve of the entire observation,
obtained by summing events from all  three XIS cameras.
The largest burst is indicated by a red arrow.
{\em Right}: A zoom  of the $\sim$3 ks of the left panel around the burst.
}
\label{fig:xis_persistent_lc}
\end{center}
\end{figure*}


\section{OBSERVATION}
\label{obs}

A ToO observation of SGR\,0501+4516 was performed with {\it Suzaku} (Mitsuda et al. 2007), 
on 2008 August 26 starting at 00:05 (UT), until August 27 08:25 (UT). 
The X-ray Imaging Spectrometer (XIS; Koyama et al. 2007)
was operated in the normal mode with 1/4 window option, 
to ensure a time resolution of 2\,s.  
The Hard X-ray Detector (HXD; Takahashi et al. 2007) 
was in the standard mode, 
wherein individual events have a time resolution of $61~\mu$s, 
and the four-channel HXD-WAM counts are available every 1 s. 
The target was placed at the ``XIS nominal" position. 

The XIS and HXD data were both processed with 
the {\it Suzaku} pipeline processing ver.\,2.2.8.20.  
Events were discarded if they were acquired in
the South Atlantic Anomaly, or in regions of low cutoff rigidity 
($\leq 6$ GV for XIS and $\leq 8$ GV  for HXD), 
or with low Earth elevation angles.  
The net exposures obtained with the XIS and the HXD were 43 ks and 55 ks,
respectively.

\begin{figure}
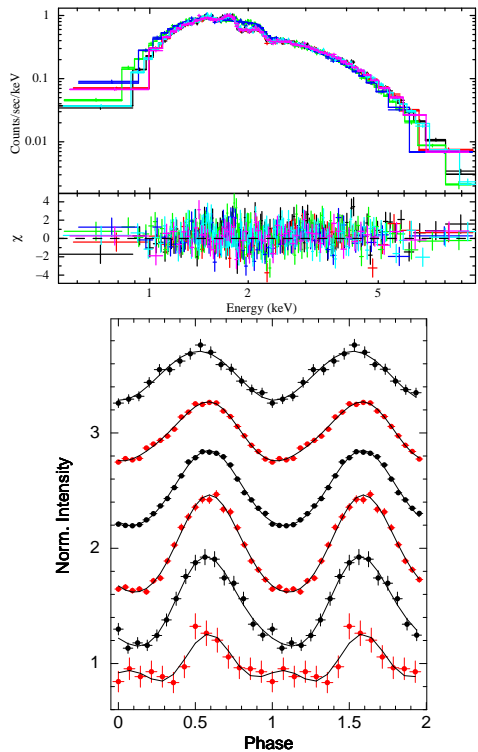

\begin{center}
\vbox{
\includegraphics[scale=0.26,angle=270]{fig2a.ps}
\includegraphics[scale=0.35,angle=270]{fig2b.ps}}
\caption{XIS results on the persistent emission from SGR\,0501+4516. 
{\em Top panel}: All XIS spectra of the persistent
emission (see text for details), 
simultaneously modeled with an absorbed blackbody plus power law. 
{\em Bottom panel}: Pulse profiles
as a function of the energy band (from top to bottom: 0.5--1, 1--2,
2--3, 3--5, 5--8, and 8-12\,keV).}
\label{fig:xis_persistent_spec}
\end{center}
\end{figure}

\section{ANALYSIS AND RESULTS}
\label{results}

\subsection{The persistent soft X-ray emission}
\label{xispersistent}

The on-source XIS events were extracted 
from each of the 3 XIS cameras, 
over a region $1'.8$ in radius centered on the target position.  
The background events were derived from 
a similar region  as far away from the source as possible.

We detected SGR\,0501+4516 at a 0.4--10~keV count rate of 1.74 count~s$^{-1}$ 
with each XIS front-illuminated (FI) CCD (XIS~0  and XIS~3), 
and 1.82 count~s$^{-1}$ with the back-illuminated (BI) one (XIS~1). 
The 0.4--10 keV XIS light curve presented in 
Figure~\ref{fig:xis_persistent_lc}  reveals 32 short ($\le 2$~s) burst episodes, 
where the count rate per 2~s, 
summed over the 3 cameras, 
increased by more than $5\sigma$ above 
the persistent emission (plus background). 
Several of them were also detected with the HXD (\S \ref{hxdburst_lc}).

After eliminating the 32 bursts, 
we began the timing analysis (using {\tt Xronos}~5.21) 
by converting all the event arrival times to the Solar System barycenter.  
A power spectrum analysis confirmed the reported 
$5.76$\,s periodicity (G\"{o}\u{g}\"{u}\c{s} et al. 2008) in several harmonics.  
By folding the data at the fundamental period 
and employing a phase-fitting technique (Dall'Osso et al. 2003), 
the best-fit period was found to be $P=5.762072\pm0.000002$~s 
at the 14704.0~TJD epoch (all uncertainties are at the 90\% confidence level).  
This is consistent with those measured with {\it RXTE}, {\it Swift}, 
and {\it XMM-Newton} (G\"{o}\u{g}\"{u}\c{s} et al.~2008; Israel et al. 2008a).

Given the XIS timing resolution of 2\,s, 
we cannot study detailed  pulse profile sub-structures. 
However, looking at the  0.5--12\,keV XIS profiles 
as a function of energy  (Figure~\ref{fig:xis_persistent_spec} bottom), 
we found that below about 5\,keV one sinusoidal function 
at the fundamental spin  frequency  fit the profile shape well, 
while at higher energies the second harmonic component
is needed at a  significance of 4$\sigma$ 
to reproduce  the profile.
The pulsed fraction (defined as the semi-amplitude of the 
fundamental sinusoidal modulation divided by 
the mean background-subtracted count rate) was 30(1)\%
on average, while it varied with energy as
21(3), 24(2), 30(2), 43(1), 30(3), and 20(3)\%,
at 0.4--1, 1--2, 2--3, 3--5, 5--7, and 7--10~keV, respectively.

Soft X-ray spectra of the persistent emission were studied with {\tt XSPEC12}.  
We modeled the background-subtracted XIS-FI (XIS 0 plus XIS 3) 
and XIS-BI spectra jointly with an absorbed blackbody plus power-law, 
a typical empirical spectral decomposition for magnetars. 
A multiplicative constant factor was used to account for calibration
uncertainties between XIS-FI and XIS-BI, which were $<3$\%.  
As presented in Figure~\ref{fig:xis_persistent_spec}, 
we found a good fit ($\chi^2_\nu=0.96$ with 603 degrees of freedom) 
with an hydrogen column density  of 
$N_{\rm H}=(0.89\pm0.08)\times10^{22}$~cm$^{-2}$, 
a photon-index of $\Gamma=2.8\pm0.1$, 
and a blackbody temperature and radius of 
$kT= 0.69\pm0.01$~keV and $3.2\pm0.2~d_{10}$\,km, respectively,
where $d_ {10}$ is the source distance in units of 10\,kpc.  
The observed and unabsorbed 0.5--10~keV fluxes were 
$3.8(1)\times10^{-11}$ and
$8.7(8)\times10^{-11}$\,erg~s$^{-1}$cm$^{-2}$, respectively. 
While a good fit was also found using 
a resonant cyclotron scattering model (Rea et al.~2008),
a two-blackbody model was unsuccessful. 
The HXD results on the persistent emission will be reported elsewhere.

\begin{figure}[bht]
\begin{center}
\hbox{
\includegraphics[scale=0.33]{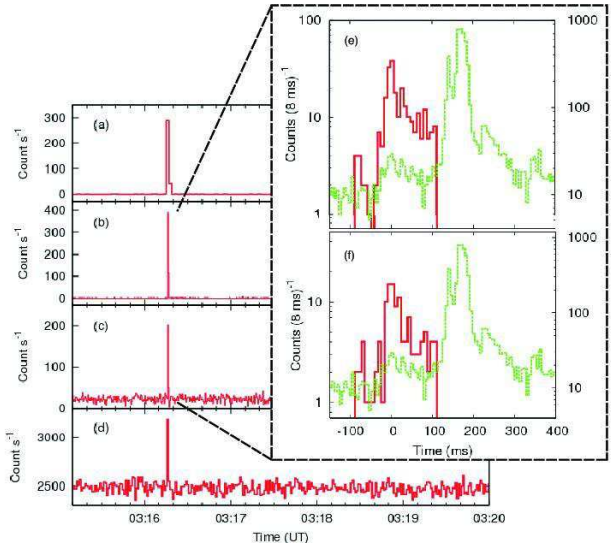}}

\caption{
Background-inclusive and dead-time-uncorrected
light curves of the largest burst.
(a) The summed 0.6--10 keV XIS data per 2 s
(\S~\ref{hxdburst_spec}).
(b), (c)  HXD-PIN (10--50 keV) and HXD-GSO (50--250 keV) data,
respectively, with 0.5 s binning.
(d) The 28--111 keV HXD-WAM data with 1 s binning.
Panels (e) and (f) are expanded views of (b) and (c), respectively, 
using 8 ms binning (left ordinate),
where the 18--1160 keV Konus-{\it Wind} light curve
(V. Palshin, private communication 2008) is 
superposed in green (right ordinate).}
\vspace*{-3mm}
\label{fig:burst_lc_fine}
\end{center}
\end{figure}

\subsection{The powerful burst and its light curves}
\label{hxdburst_lc}

Among the 32 bursts detected by the XIS (\S~\ref{xispersistent}), 
four were also detected by HXD-PIN,
and three of them by HXD-GSO, 
all with $>3\sigma$ significance. 
As shown in Figure~\ref{fig:burst_lc_fine} with a  red arrow, 
the largest event  on  2008 August 26 at 03:16:16.947 (UTC) 
was detected by all the {\it Suzaku} instruments, 
including the four HXD-WAM. 
It was also observed by the Konus-{\it Wind} instrument (Palmer  2008). 
The HXD-PIN and HXD-GSO light curves are rather similar 
with negligible relative delays (within $\pm 8$ ms), 
indicating that the spectrum is not significantly time-dependent.

The HXD-PIN signal intensity at the peak and that averaged over the
$\sim 200$\,ms duration are $\sim 38$ and $\sim 10$ counts per 8\,ms,
which translate to a source intensity of $\sim 89$ and $\sim 23$\,Crab, respectively.  
Up to  time $t\sim$107\,ms, these light curves are
relatively free  from instrumental dead times ($\lesssim 6\%$), 
which have three components (Takahashi et al. 2007); 
processing times in the analog electronics (HXD-AE), 
the limited transfer rate from HXD-AE to the digital electronics (HXD-DE), 
and that from HXD-DE to the spacecraft data processor.

At $t\sim$107\,ms, the HXD signals suddenly terminated. 
This is an instrumental effect, 
caused by the second and third factors above,
including a forced ``flush" of an HXD-DE output buffer.  
The number of ``lost" HXD events can be 
estimated from various scalar information.
Over a 4-s interval including the burst, 
we found that HXD-PIN received  $\sim 14,000$ photons, 
of which only $\sim 250$, or $\sim$2\%, 
were duly processed onboard and edited into the light curve.  
This fraction is similar in the HXD-GSO data, 2.7\% (195 processed vs. 7164 received).

By comparing these {\it Suzaku} light curves with the Konus-{\it Wind}
data (green in Fig.~\ref{fig:burst_lc_fine}e and \ref{fig:burst_lc_fine}f), 
we found that the time history observed
by the HXD is a ``precursor" event.  
When the much stronger main burst began at $t\sim 138$ ms, 
the HXD data became completely suppressed, 
until they returned to normal at $t=740$ ms.  
This is consistent with the main burst duration, $\lesssim 0.5$ s, 
recorded by Konus-{\it Wind}.  
The entire burst  thus contained $\sim 50$ times more  photons 
in the hard X-ray band than actually detected by the HXD.

In the XIS data, this burst was split into two readout frames of 2\,s each.  
We accumulate the XIS data over the 4\,s, which contains not
only the precursor but also the large main peak (missed by the HXD).
Around the XIS image centroid, 
the very high burst intensity caused strong event pile-up, 
and telemetry saturation 
which in turn caused some truncation of the XIS frame readout.


\begin{figure}[htbp]
\begin{center}
\includegraphics[scale=0.3]{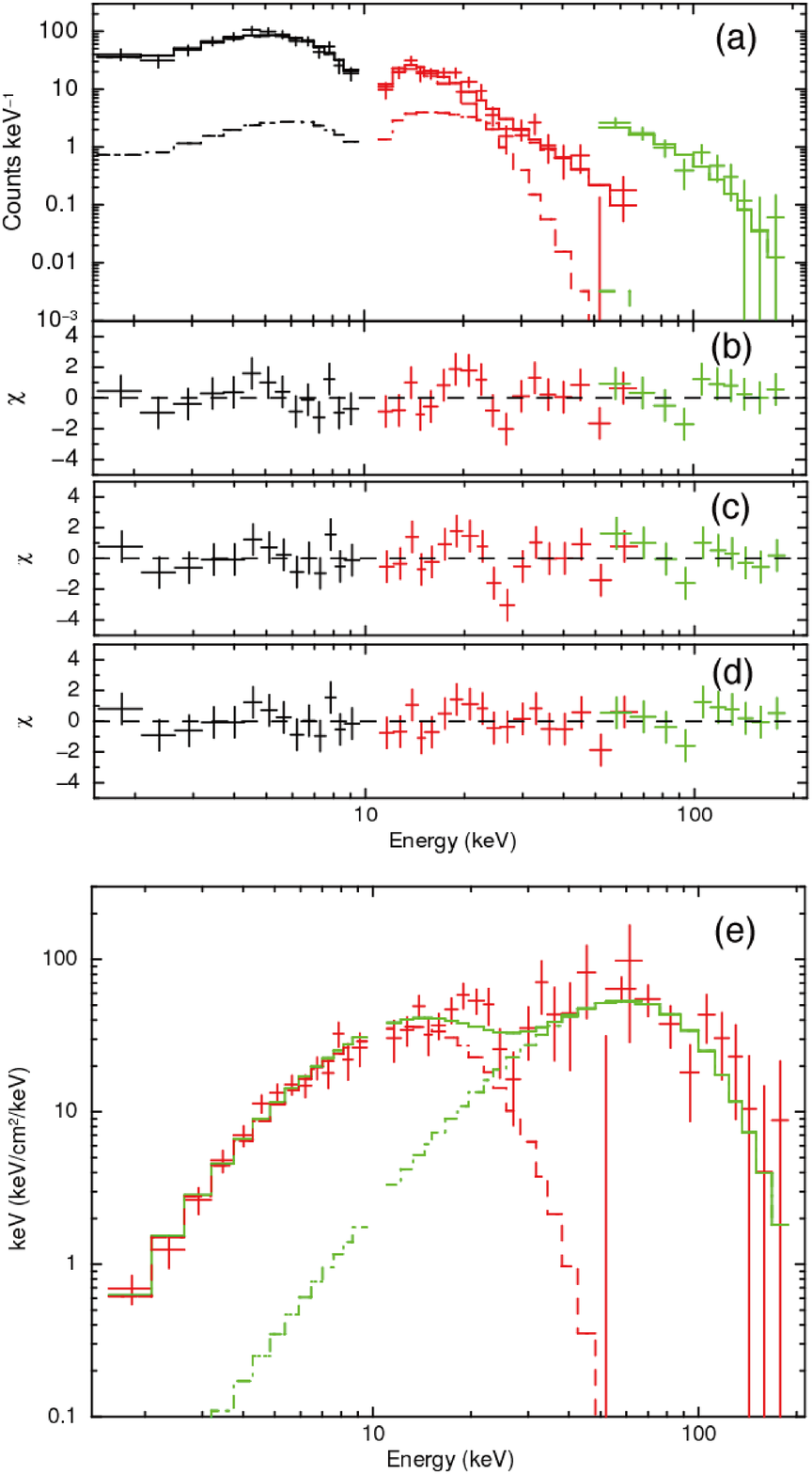}

\caption{ (a) Background-subtracted broad-band spectra of the largest burst, 
fitted by a two blackbody model. Black, red, and green specify
XIS-FI, HXD-PIN, and HXD-GSO, respectively.  
The ordinate is not divided by exposure.  
(b) Residuals from the fit in panel a.  
(c)(d) The same as panel b, but when using {\tt CompTT} 
and {\tt CompTT $\times$ cyclabs} models, respectively.
(e) The $\nu F_{\nu}$ plot corresponding to panel a.
The XIS data points are scaled to 6\% of those in panel (a).}
\label{fig:burst_spectra}
\end{center}
\end{figure}


\vspace*{-5mm}
\subsection{The burst spectra}
\label{hxdburst_spec}

Figure~\ref{fig:burst_spectra}a shows the spectra of the largest burst, 
where the ordinate is counts per unit energy (not divided by exposure).  
The HXD data were accumulated over a time interval of $\sim200$\,ms 
around the {burst arrival time (see Figure~\ref{fig:burst_lc_fine}). 
We subtracted the HXD-GSO background (22 count s$^{-1}$, 50-250 keV) 
using a 400\,s interval before and after the burst.  
The HXD-PIN background (0.56\,count\,s$^{-1}$, 10--70\,keV) was negligible.  
Thus, the emission is detectable up to $\sim 200$\,keV.

The XIS spectra in {Figure~\ref{fig:burst_spectra}a were accumulated
from XIS-FI over the two frames containing the burst. 
Unlike the HXD data which cover only the precursor, 
the XIS events must be dominated by the main burst.  
To avoid the pile-up and readout truncation problems,
we extracted the XIS~0 events using an annulus between radii
of $2'$ and $4'.5$, and XIS~3 events between $3'$ and $4'.5$.  
The two XIS-FI cameras thus yielded 515 burst photons, 
among which pile-up events are estimated as $<$20\%.  
We generated particular XIS ancillary response files 
that correctly reproduce the fraction of photons 
falling onto these limited accumulation regions.

To evaluate the precursor spectrum, we jointly fitted the HXD-PIN
(10--70 keV) and the HXD-GSO (50--250 keV) spectra.  
The 1--10 keV XIS-FI data were also incorporated, 
assuming that the precursor has nearly the same spectrum 
as the much brighter main burst.  
To take into account the various uncertainties  in the HXD and XIS
data processing, 
we left relative model normalizations free between
them.  The interstellar absorption was fixed at $0.89 \times 10^{22}$
cm$^{-2}$ as specified by the persistent emission.

As shown in Figure~\ref{fig:burst_spectra}, 
a model comprising two blackbodies 
(typical for magnetars bursts; e.g. Olive et a. 2004; Feroci et al.~2004;
Nakagawa et al. 2007; Israel et al.~2008b; Esposito et al.~2008)
reproduced the data fairly well with $\chi_{\nu}=41.0/35=1.17$.  
The deconvolved $\nu F_\nu$ spectra are shown in Figure~\ref{fig:burst_spectra}e.  
The fit yielded two temperatures:
$3.3_{-0.4}^{+0.5}$\,keV and $15.1_{-1.9}^{+2.5}$\,keV.  
Assuming spherical emission regions, 
the the cooler and hotter blackbodies are estimated to have radii of 
$8.0_{-2.1}^{+2.9}~d_{10}$\,km 
and $0.46_{-0.14}^{+0.16}~d_{10}$\,km, respectively.  
The average (over the 200 ms) and peak 1--200 keV fluxes of the precursor 
are $1.0\times 10^{-6}$ and $3.8\times 10^{-6}$\,erg\,s$^{-1}$\,cm$^{-2}$, 
respectively, 
and the corresponding fluence is $2.0\times 10^{-7}$\,erg\,cm$^{-2}$.  
The total burst fluence is estimated to be $\sim 50$ times higher. 
The model normalization for the HXD data (PIN and GSO together)
was  $\sim 6\%$ of that of XIS-FI.

When a thermal Comptonization model, {\tt CompTT},  is alternatively used,
the fit becomes slightly worse ($\chi_{\nu}=42.8/35=1.22$; 
Fig.~\ref{fig:burst_spectra}c).

Although the above models  are roughly successful for  the burst spectrum, 
a hint of negative residuals is present
at $\sim 26$\,keV (Fig.~\ref{fig:burst_spectra}b),
where the two blackbodies cross over.
Applying  a Gaussian absorption ({\tt Gabs})
or cyclotron absorption ({\tt Cyclabs}) factor at $\sim 26$ keV,
the fit was improved to $\chi_\nu =0.89$ (Fig.~\ref{fig:burst_spectra}d).
However, its  addition is  significant only at 1.9$\sigma$ level.

\section{DISCUSSION}
\label{discussion}

SGR\,0501+4516 was observed by {\it Suzaku} four days after the burst
activation at a flux level of 
$\sim 3.8\times10^{-11} \ {\rm  erg\,s^{-1}\,cm}^{-2}$ (0.5--10 keV).
The persistent 0.5--10 keV spectrum obtained with the XIS is 
described by  a $kT= 0.69$~keV blackbody 
plus a  $\Gamma \sim 2.8$ power-law.

If the identification of SGR\,0501+4516 with the {\em ROSAT}  source 
2RXP J050107.7+451637 (Kennea \& Mangano 2008) is correct,
the unabsorbed 0.2--10 keV flux increased by a factor $\sim 15$ 
with respect to the 1992 value (for a power-law  model). 
Large spectral and flux variations have been recorded in other SGRs as well, 
in coincidence with the onset of phases of  bursting activity. 
The two most active repeaters, SGR\,1806$-$20 and, 
to a  lesser extent, SGR\,1900+14, are known to undergo 
rather gradual variations  in luminosity and spectral properties.
In fact, prior to the Giant Flare of 2004 December 27th 
(Hurley et  al. 2005; Palmer et al. 2005), 
SGR\,1806$-$20 exhibited a factor  $\sim 2$  flux increase and a decrease in 
$\Gamma$  from $\sim 2$ to $\sim 1.6$ (Mereghetti et al. 2005).
In the case of  SGR\,1900+14\, in 2006,
the spectral index went from $\sim 2.4$  to $\sim 2$ 
and the flux increased by  $\sim 40\%$ (Israel et al. 2008b).
In contrast,
sources which  experience long stretches of quiescence 
show much more dramatic  changes.  
In May 2008, when SGR\,1627$-$41 reactivated after more than 10  yrs, 
its $\Gamma$ switched from $\sim 3.3$ to $\sim 1.5$, 
and  the observed 2--10 keV flux increased 
by a factor of 40 (Esposito et al. 2008). 
While the luminosity increase in SGR\,0501+4516 
is not dissimilar to that of SGR\,1627$-$41,
while in outburst its persistent spectrum is much softer.

Among other short X-ray bursts from SGR\,0501+4516,
we detected a very powerful one,
observed also by Konus-{\it Wind} (V. Palshin private communication 2008).
The 1--200 keV precursor spectrum 
detected with the XIS and the HXD was fitted 
reasonably well by the two blackbody model. 
One possible interpretation of the two temperatures
($\sim$3 keV and $\sim$15 keV) is 
that they represent the photospheres of the ordinary  and extraordinary modes 
(Harding \& Lai 2006; Israel et al. 2008b), 
which can have different radii and temperatures due to
suppression of extraordinary-mode scattering 
and photon-splitting in the super-critical magnetic field.

\acknowledgements We thank Dr. Valentin Pal'shin and Dr. Dmitry
Frederiks for allowing us to use their Konus-{\it Wind} light curve
prior to publication.  Our thanks are also due to the {\it Suzaku}
operation team, who successfully conducted the ToO observation. NR is
supported by an NWO Veni Fellowship, DG thanks the CNES for financial
support, and SZ acknowledges support from STFC.

\end{document}